\documentclass[12pt]{iopart}
\usepackage{graphicx}
\usepackage{color}
\usepackage{hyperref}
\usepackage{ulem}

\begin{document}
\title[]{Geometrical analysis of Kerr-lens mode-locking for high-peak-power ultrafast oscillators}

\author{Peiyu Xia$^1$, Makoto Kuwata-Gonokami$^1$, Kosuke Yoshioka$^{1,2^{*}}$}

\address{$^1$Department of Physics, Graduate School of Science, The University of Tokyo, 7-3-1 Hongo, Bunkyo-ku, Tokyo 113-0033, Japan}
\address{$^2$Photon Science Center, School of Engineering, The University of Tokyo, 2-11-16 Yayoi, Bunkyo-ku, Tokyo 113-8656, Japan}

\ead{$^{*}$yoshioka@fs.t.u-tokyo.ac.jp}
\vspace{10pt}
\begin{indented}
\item[]March 2020
\end{indented}

\begin{abstract}
Ultrashort pulses from Kerr-lens mode-locked oscillators have inspired a variety of applications.
The design and alignment of these laser resonators have thus far been theoretically supported by the conventional analysis of beam propagation.
However, the well-established theoretical framework is sometimes beyond the scope of high-peak-power oscillators.
%
In this paper, we analyze the geometry of ring resonators by extending the ABCD-matrix method to a high-peak-power regime.
The guidelines to achieving stable Kerr-lens mode-locking is provided for high-peak-power pulses. 
\end{abstract}
%
%
%
%
%
\section{Introduction}
{Following the demonstration of Kerr-lens mode-locking (KLM) by using the optical Kerr effect in Ti:Sapphire \cite{Spence1991}, laser resonators with a Kerr medium have become the basic platform for the generation of ultrashort optical pulses.
Besides, the emergence of Ti-, Yb-, and Cr-doped gain crystals allows to generate femtosecond pulses in the visible to the near-infrared region \cite{Evans1992,Sennaroglu1994,Petrov1997,Cizmeciyan2009}. 
In designing these Kerr-lens mode-locked oscillators, one of the essential tasks is to judge if the laser can establish stable KLM.

To evaluate the stable KLM operation quantitatively, earlier theoretical works have presented a basic framework described as the following \cite{Brabec1992,Magni1993}.
First, KLM is caused by the power-dependent gains or losses induced by the self-focusing effect in a Kerr medium.
Second, the slope of the gains or losses is evaluated by numerically analyzing the beam propagation in the spatial domain.
Here, the essential parameter to evaluate the slope is the relative spot-size variation defined as follows:
\begin{equation}
\label{Deltafunc1}
\eqalign{
\delta=\left[-\frac{1}{\omega} \frac{d \omega}{d P}\right]_{P=0}.
}
\end{equation}
$\omega(P)$ is the spot size at the Kerr medium as a function of the intracavity power, $P$.
So far, the $\delta$ parameter has been mostly used as a measure of stability to provide the basic guidelines for KLM in the commonly-used linear or ring resonators, and to design these resonators \cite{Magni1993v2,Cerullo1994,Magni1996,Georgiev1992,Brabec1993,Heatley1993,Agnesi1994}.
Although the more realistic frameworks which treat the spatiotemporal dynamics have been developed \cite{Jirauschek2003,Wright2020}, the analysis based on the $\delta$ parameter is still widely used \cite{Donin2012,Yefet2013,Klenner2015,Li2017,Sugiyama2018} due to its simplicity and low cost of computation.

However, the $\delta$ parameter is not always valid when the intracavity power $P$ is close to or larger than the critical power for self-focusing, $P_\mathrm{c}$, defined as
\begin{equation}
\label{CP}
P_{\mathrm{c}}=\frac{\alpha \lambda^{2}}{4 \pi n_{0} n_{2}},
\end{equation}
where $n_0$ and $n_2$ are the linear and nonlinear refractive indices of the Kerr medium, $\lambda$ is a laser wavelength, and $\alpha$ is a constant calculated to be 1.86–2.0 for Gaussian beam \cite{Magni1993,Fibich2000}.
The theoretical work by Larotonda showed that the stability of KLM saturates when $P/P_\mathrm{c}$ approaches 1 in a Kerr medium \cite{Larotonda2003}.
Besides, Bartels and Kurz pointed out that experimentally realized oscillators including their result usually satisfied $P/P_\mathrm{c}>1$ \cite{Bartels2002}.
For instance, the Ti:Sapphire resonator of Ref. \cite{Spence1991} satisfied $P/P_\mathrm{c}=1.3$, calculated from the reported output peak power of 90 kW with a 3.5\% output coupler and $P_\mathrm{c}= 2$ MW for Ti:Sapphire.
In the same way, the Cr$^{4+}$:YAG resonator of Ref. \cite{Sennaroglu1994} satisfied $P/P_\mathrm{c}=1.4$, and the Cr$^{2+}$:ZnSe resonator of Ref. \cite{Cizmeciyan2009} satisfied $P/P_\mathrm{c}=1.2$.
We note that these works are the first demonstration of KLM using the corresponding gain crystals.
Therefore, it is practically necessary to understand the mechanism of stable KLM in laser resonators with $P/P_\mathrm{c}>1$.

In this work, we numerically analyze beam propagation in ring resonators particularly in the high-peak-power regime ($P/P_\mathrm{c}$ up to 1.6).
Based on the ABCD-matrix formulation, we find that the stable region of the ring resonator for the pulse propagation differed significantly from that for the continuous-wave (CW) propagation which corresponds to $P/P_\mathrm{c}=0$.
Moreover, we extend the analysis based on the $\delta$ parameter to better evaluate the stability of KLM. 
}
\section{Method}
\begin{figure}[t]
\begin{center}
\includegraphics[width=110mm]{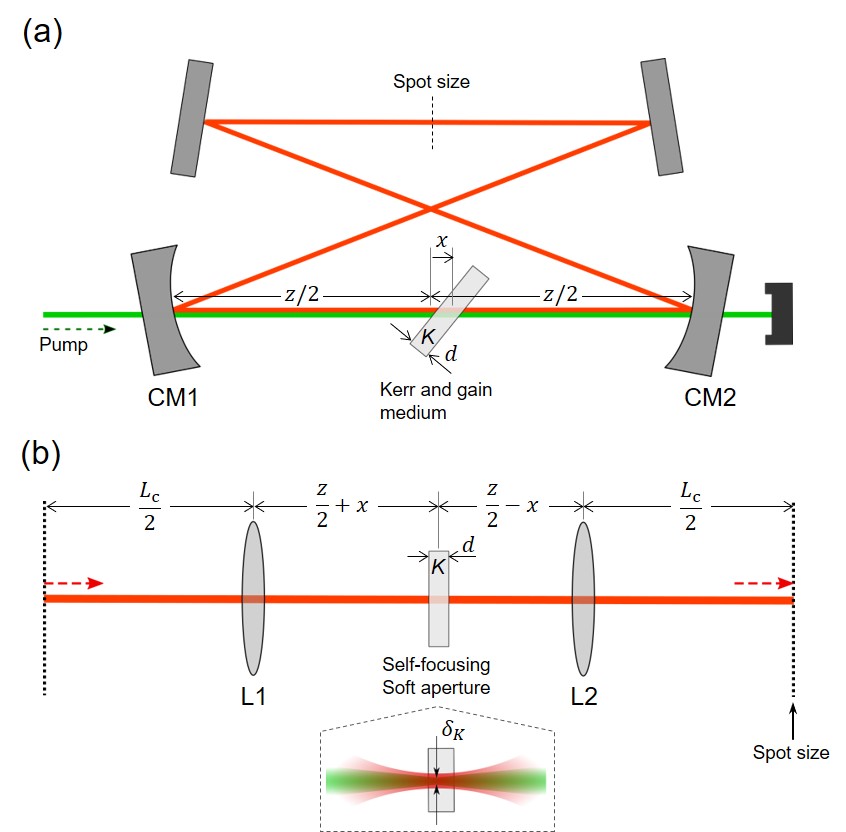}
\caption{(a) Configuration of a typical Kerr-lens mode-locked ring oscillator, and (b) the representative model for the ABCD-matrix calculation. The black dotted line represents the periodic propagation of a circulating beam and corresponds to the fixed position for the calculation of the spot size. $x$ is the displacement of the medium from the center position, and $d$ is the thickness of the medium, neglecting the oblique incidences,  $z$ and $L_\mathrm{c}$, are the shorter and longer propagation distances between the concave mirrors, CM1 and 2, respectively. The inset illustrates the assumption of the pump-to-laser overlap and the position to calculate $\delta_K$. L1 and 2: a pair of lenses with the same focal length as CM1 and 2.
}
\label{fig1}
\end{center}
\end{figure}
As shown in Fig. \ref{fig1}, we analyze the ring resonator with neglect of the astigmatism of oblique incidence to a Kerr medium and mirrors.
Consequently, the beam propagation in the ring resonator is equally represented as in Fig. \ref{fig1}(b).
We assume that the spot size of the pump beam is always aligned smaller than that of the resonant intracavity beam, as shown in the inset of Fig. \ref{fig1}(b).
This alignment is a common technique called “soft-aperture KLM” \cite{Brabec1992}, where the higher gain favors the smaller spot size in the gain medium.
The radius of curvature of each concave mirror is 30 mm, corresponding to the focal length, $f$, of 15 mm.
The thickness, $d$, and refractive index, $n_0$, of the medium is 2.5 mm and 1.76, respectively. $L_\mathrm{c}$ is 180 mm, and the laser wavelength, $\lambda$, is 800 nm.
These are the conventional parameters used for KLM in the Ti:Sapphire ring resonators with GHz-level repetition rates \cite{Bartels2002,Bartels1999}, {which are also estimated to satisfy $P/P_\mathrm{c}>1$.}
Compared with the linear resonator, the ring resonator has a single stable region, as described in Eq. \ref{CWstr}, whereas the linear resonator has double stable regions \cite{Brabec1992}.
Therefore, the understanding of KLM, based on the ring resonator, is clearer and analogous to the linear resonator.

We first use the formalism based on the ABCD matrix to calculate the self-consistent Gaussian mode.
In the case of the CW propagation, the stable region is given by
\begin{equation}
\label{CWstr}
2 f+d-d_{\mathrm{e}}<z<\frac{2 f L_{\mathrm{c}}}{L_{\mathrm{c}}-2 f}+d-d_{\mathrm{e}},
\end{equation}
where $d_{\mathrm{e}}=d/n_0$ is the effective length of the medium, and $z$ is the distance between the concave mirrors.
In our case, Eq. \ref{CWstr} corresponds to $31.08\ \mathrm{mm}<z<37.08\ \mathrm{mm}$.
These lower and upper bounds are defined as $z_\mathrm{min}$ and $z_\mathrm{max}$ repectively and indicated in Fig. 2.
In the case of the pulse propagation, nonlinear propagation in the Kerr medium caused by self-focusing is expressed by the following matrix \cite{Magni1993}.
\begin{equation}
\label{KerrMatrix}
\mathbf{M}(\gamma)=\sqrt{1-\gamma}\left(\begin{array}{cc}
1 & d_{\mathrm{e}} \\
-\frac{\gamma}{(1-\gamma) d_{\mathrm{e}}} & 1
\end{array}\right).
\end{equation}
The $\gamma$ parameter depends on the intracavity power and is expressed by
\begin{equation}
\label{gamma}
\gamma=K\left[1+\frac{1}{4}\left(\frac{2 \pi \omega_{\mathrm{c}}^{2}}{\lambda d_{\mathrm{e}}}-\frac{\lambda d_{\mathrm{e}}}{2 \pi \omega_{0}^{2}}\right)^{2}\right]^{-1},
\end{equation}
where $K=P/P_\mathrm{c}$ is the normalized intracavity power, $\omega_{\mathrm{c}}$ is the spot size at the center of the medium, and $\omega_{0}$ is the spot size at the beam waist calculated as if $K=0$.
The formalism using both linear ABCD matrices and the matrix of Eq. \ref{KerrMatrix} allows to calculate the spot size for the pulse propagation.
However, Eq. \ref{gamma} has difficulty with multivariable analysis, because both $\omega_{\mathrm{c}}$ and $\omega_{0}$ depend on the resonator parameters and the intracavity peak power.
For multivariable analysis, we modify Eq. \ref{gamma} as follows.
\begin{equation}
\label{gamma2}
\gamma=K\left[1+\left( \frac{d^{-1}+\mathrm{Re}\left(q^{-1}\right)}{\mathrm{Im}\left(q^{-1}\right)} \right)^{2}\right]^{-1}.
\end{equation}
where $q^{-1}=R_{\mathrm{s}}^{-1}-i \lambda / \pi n_{0} \omega_{\mathrm{s}}^{2}$ is the reciprocal of $q$, given by the radius of the curvature, $R_\mathrm{s}$, and the spot size, $\omega_\mathrm{s}$, at the incident surface of the medium.
After the parameters are set, the calculation of the round-trip ABCD matrix of the resonator, including matrix $\mathbf{M}(\gamma)$, leads us to find the $q$ parameter of any plane, which becomes the iterative solution to the self-consistent equation, $q=[A(\gamma) q+B(\gamma)] /[C(\gamma) q+D(\gamma)]$, with $\gamma=\gamma(q)$ given by Eq. \ref{gamma2}.

Next, we define the following $\delta_K$ parameter, extended from Eq. \ref{Deltafunc1}, to evaluate the stability of KLM.
\begin{equation}
\label{deltaK}
\delta_{K}=\left[-\frac{1}{\omega} \frac{d \omega}{d\left(P / P_{\mathrm{c}}\right)}\right]_{P=K P_{\mathrm{c}}}.
\end{equation}
Eq. \ref{deltaK} represents the relative spot size variation in perturbation to the normalized peak power of $K$, i.e. the robustness against the spatial perturbation.
The case of $K=0$ corresponds to Eq. \ref{Deltafunc1}.
The negatively larger $\delta_K$ in the gain medium leads to a higher gain under the soft-aperture condition as represented in the inset of Fig. \ref{fig1}(b).
Compared with Eq. \ref{Deltafunc1}, the $\delta_K$ parameter deviates from the $\delta$ parameter when $K$ approaches 1.
Therefore, the evaluation of $\delta_K$ is more desirable to achieve intracavity peak power around $P_{\mathrm{c}}$ during the KLM process.
For example, if $\delta_K$ is positive at certain peak power, the pulse propagation is unstable due to no gain or losses, which results in a decrease of the peak power.
This gain-loss balance also causes the mode-locking process in the time domain, as described in the role of saturable absorbers and self-amplitude modulation \cite{Brabec1991,Kartner1998,Haus2000}.
\section{Result}
\begin{figure}[t]
\begin{center}
\includegraphics[width=150mm]{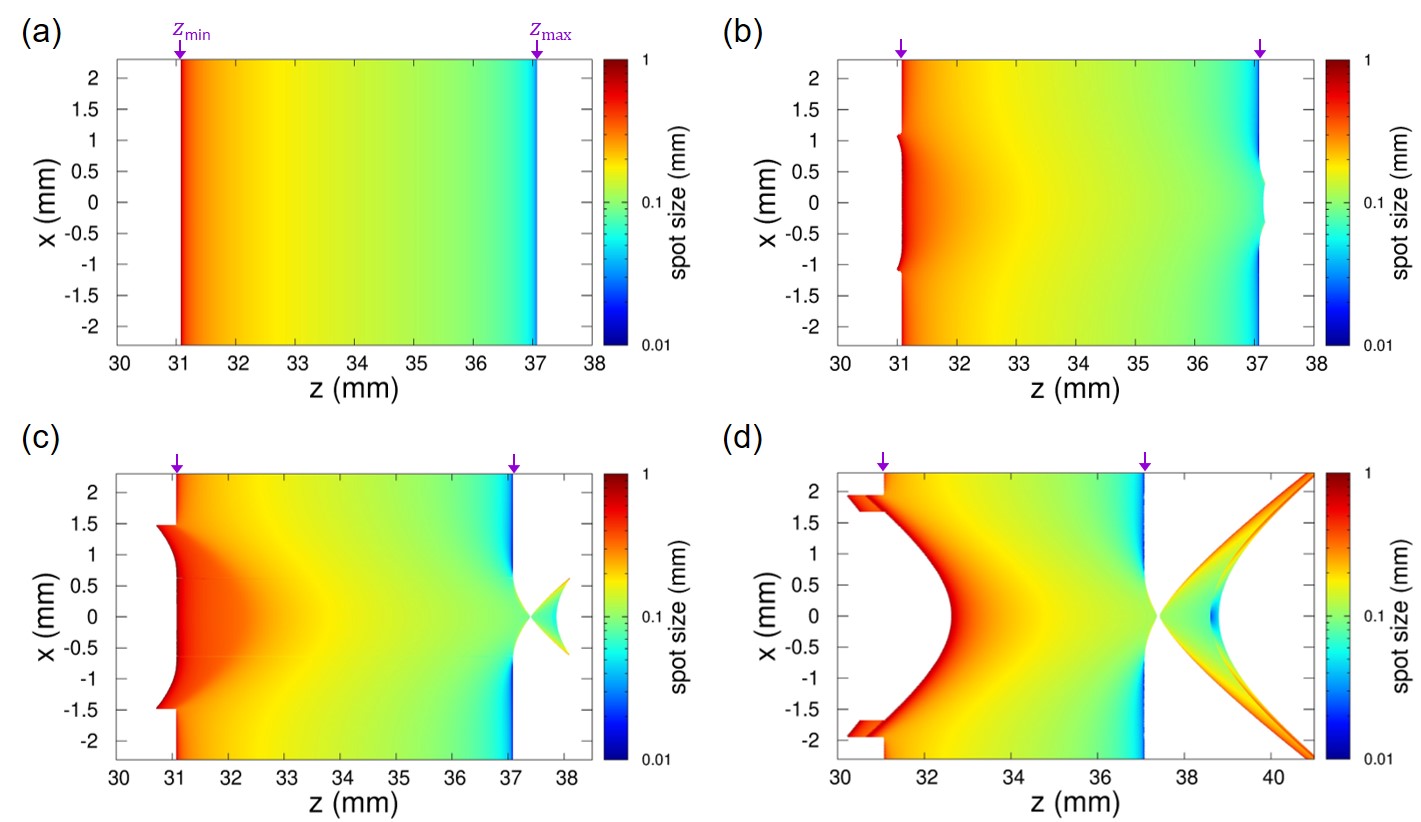}
\caption{Variation of the spot size at the fixed position as indicated in Fig. \ref{fig1}(b). The normalized intracavity peak power, $K$, is (a) 0.0, (b) 0.4, (c) 0.8, and (d) 1.6, respectively. The purple arrows indicate the inner and outer stability edges for the CW propagation expressed by Eq. \ref{CWstr}.}
\label{fig2}
\end{center}
\end{figure}
%
Figure \ref{fig2} shows the variation of the spot size as a function of $z$ and $x$, the position of which is fixed in the resonator (see Fig. \ref{fig1}(b)). 
We found the self-consistent solution only in the colored region, where the calculated spot size is finite.
The finite value out of the color range is also displayed with the maximum or minimum color.
In some conditions near the stability edge, we found two self-consistent solutions, as reported on the linear resonator \cite{Magni1993}.
However, one is always discontinuous as a function of $z$ and $x$ and is therefore excluded.
Figure \ref{fig2}(a) shows that the spot size for the CW propagation ($K=0$) is independent on the medium position $x$ at any $z$.
On the other hand, significant $x$-dependencies are observed in Figs. \ref{fig2}(b)–(d).
When the intracavity power is $0.4P_{\mathrm{c}}$ as in Fig. \ref{fig2}(b), the spot-size variation changes significantly from that for the CW propagation, especially near the stability edge ($z\simeq z_{\mathrm{min}}$ and $z\simeq z_{\mathrm{max}}$).
When the intracavity power is higher than $P_{\mathrm{c}}$ as is the case of Fig. \ref{fig2}(d), parts of the self-consistent solution are no longer found near the inner stability edge with the medium close to the center position ($z\simeq z_{\mathrm{min}}$ and $x\simeq 0$).
This feature indicates that the initially stable CW propagation can evolve into the unstable pulse propagation during the KLM process.
The contrasting feature in Figs. \ref{fig2}(b)-(d) is the appearance of two pairs of sharp edges in the stable region with increasing the intracavity power.
The sharp edges grow into the unstable region for the CW propagation ($z< z_{\mathrm{min}}$ or $z_{\mathrm{max}}<z$), as can be seen from Fig. \ref{fig2}(a) to Fig. \ref{fig2}(d).
This behaviour indicates that the stable region for the pulse propagation can extend well beyond that for the CW propagation, and finally into a fish-like shape.
Therefore, the initially unstable CW propagation can be developed into the stable pulse propagation, provided that the concave mirrors and the medium position, $z$ and $x$, are properly aligned.
{These differences of the stable region between the pulse and CW propagation provide insight into some of the experimental results that showed the increase of the laser power by an order of magnitude during the KLM process, such as 1.2 W from 20–40 mW \cite{Bartels2002}, and 900 mW from $\sim$50 mW \cite{Bartels1999}.}

\begin{figure}[t]
\begin{center}
\includegraphics[width=150mm]{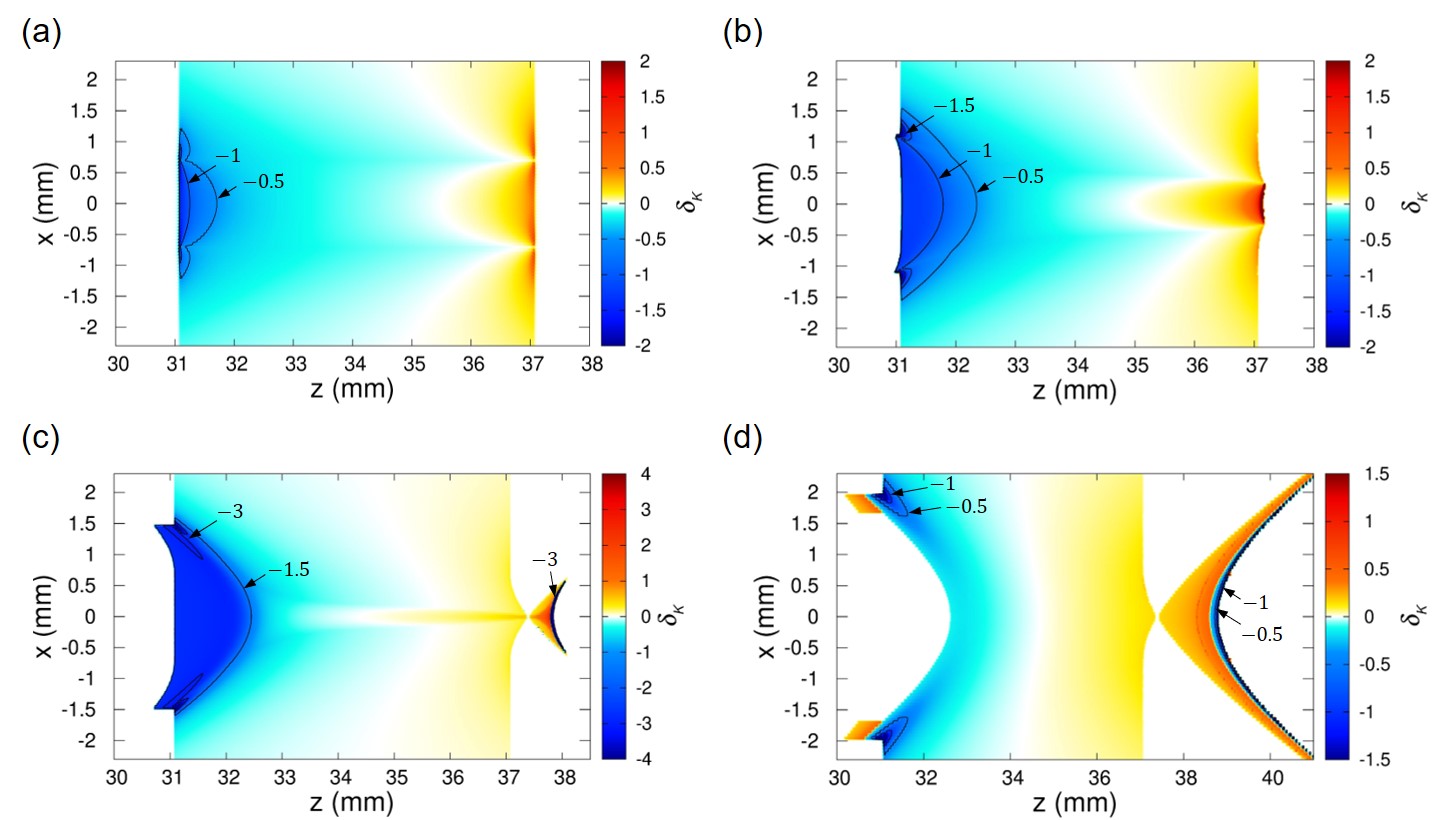}
\caption{Variation of the stability parameter, $\delta_K$. The normalized intracavity peak power, K, is (a) 0.0, (b) 0.4, (c) 0.8, and (d) 1.6, respectively. The contour lines enclose the areas having negatively large values.}
\label{fig3}
\end{center}
\end{figure}
Next, we evaluate the pulse stability to find the optimal condition to achieve stable KLM in the high-peak-power regime.
Figure \ref{fig3} shows the variation of $\delta_K$ as a function of the same resonator parameters as used in Fig. \ref{fig2}.
In each of the resonator conditions, the $\delta_K$ parameters are calculated at the position of the minimum spot size inside the gain medium, as indicated in the inset of Fig. \ref{fig1}(b).
When the intracavity power is much lower than $P_{\mathrm{c}}$ ($K\ll 1$) which almost corresponds to Fig. \ref{fig3}(a), the stability parameter is almost the same as given by Eq. \ref{Deltafunc1}.
The negatively large value can be found near the inner stability edge with the medium close to the center position ($z\simeq z_{\mathrm{min}}$ and $x\simeq 0$).
This condition of alignment is consistent with the experimental condition needed to achieve soft-aperture KLM of the ring resonators [30,31].
In contrast, the high-peak-power pulses do not favor this condition, because the self-consistent solution will disappear with increasing the intracavity power (see from Fig. \ref{fig2}(a) to Fig. \ref{fig2}(d)).
Instead of this unfavorable condition, if the medium is displaced from the center position to some extent, the self-consistent solution can be found near the sharp edges (ex. $1.5\ \mathrm{mm}\leq|x|$ for $K=1.6$ as in Fig. \ref{fig2}(d)).
Another item of note is the condition where $\delta_K$ has a negatively large value as mapped in Fig. \ref{fig3}.
The negatively large area (deep blue area) is regarded as optimal to obtain the pulses with the respective peak power, because the negatively large value corresponds to the amount of the pulse stability.
These areas are mainly found at $z\simeq z_{\mathrm{min}}$ and $|x|\simeq1.2,\ 1.5,\ \mathrm{and}\ 2.0$ mm for Figs. \ref{fig3}(b)–(d), respectively.
Therefore, the proper displacement of the Kerr medium according to the intracavity power is essential to achieve stable KLM.

We note that the other negatively large areas are also found in the unstable region for the CW propagation ($z< z_{\mathrm{min}}$ or $z_{\mathrm{max}}<z$).
These conditions may allow pulses to be built up during the KLM process by a strong mechanical perturbation to a laser resonator, or by an external injection seeding.

\begin{figure}[t]
\begin{center}
\includegraphics[width=110mm]{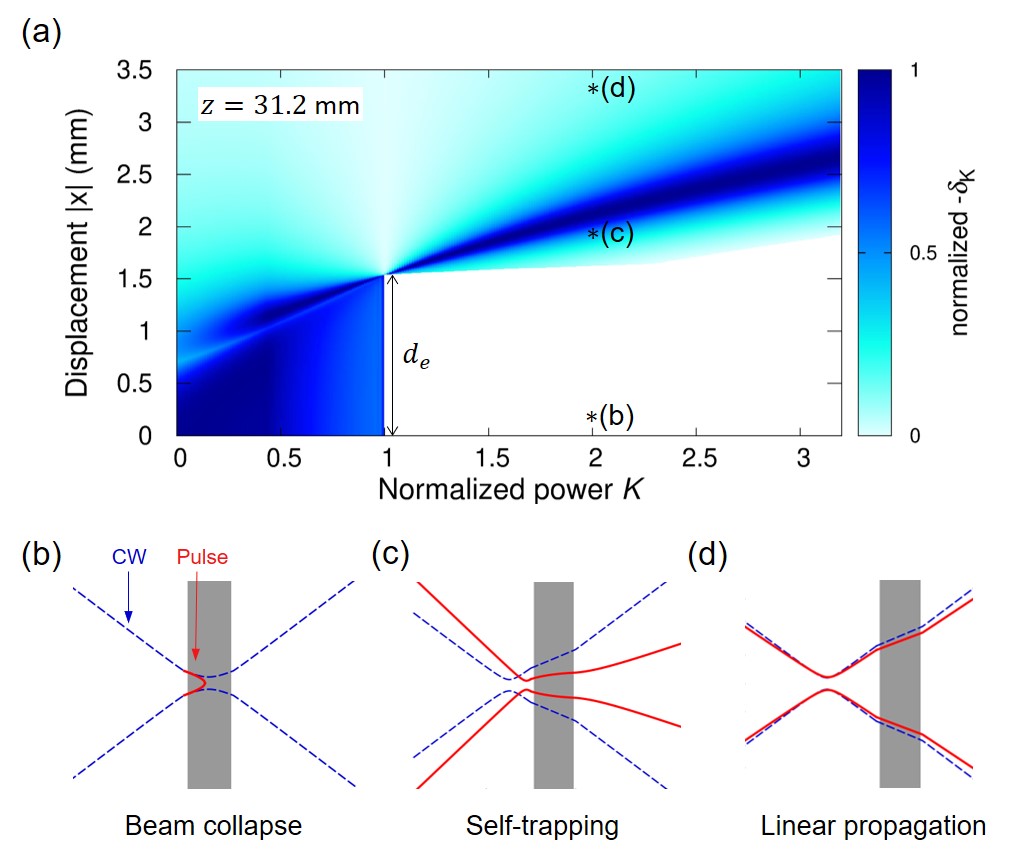}
\caption{(a) Power scaling in soft-aperture KLM with respect to the medium displacement. The stability parameter $-\delta_K (x)$ is normalized to 1 for each value of $K$. (b)-(d) The pulse propagation (red lines) around the medium is simulated on the respective conditions as marked in (a). The CW propagation (blue dashed lines) is shown at the same displacement assuming linear propagation (K=0).}
\label{fig4}
\end{center}
\end{figure}
To understand the dynamics of KLM with respect to the medium position, figure \ref{fig4} shows the three types of behaviour in the pulse propagation caused by the strong self-focusing effect inside the medium.
As shown in Fig. \ref{fig4}(a), we fix the $z$ parameter to 31.2 mm, which is close to the stability edge ($z_\mathrm{min}$), and analyze the pulse propagation with the fixed power of $2.0P_\mathrm{c}$. 
First, figure. \ref{fig4}(b) indicates that the beam with its focus position inside the medium causes the self-focusing collapse.
This condition is satisfied almost when the displacement of the medium is shorter than the effective thickness, $d_\mathrm{e}=d/n_0$, 
The collapse of the beam could not result in stable KLM, due to damage to the medium or pulse breakup.
Second, when the displacement of the medium is near the optimal area (deep blue area), the focus position of the beam is found to be usually close to the surface of the medium.
In this condition as shown in Fig. \ref{fig4}(c), the counterbalance between linear diffraction and self-focusing results in the effective self-trapping effect.
This behaviour suggests that the self-trapping effect contributes significantly to the stable KLM in the high-peak-power regime, and moreover to the appearance of the sharp edges in Fig. \ref{fig2}(b)–(d).
At last, the further displacement of the medium result in an approach to the CW mode as shown in Fig. \ref{fig4}(d).
Although the pulse propagation may be stable due to the relatively small but negative value of $\delta_K$, the condition of Fig. \ref{fig4}(c) may as well be employed. 

To summarize the discussion in Figs. \ref{fig2}–\ref{fig4}, the following guidelines are provided to achieve soft-aperture KLM in ring resonators.
\begin{itemize}
\item The alignment condition in the CW operation should be close to the inner stability edge calculated by Eq. \ref{CWstr}.
\item When the Kerr medium is aligned at the center of the concave mirrors, KLM can be only achieved when the intracavity peak power is much lower than the critical power calculated by Eq. \ref{CP}.
\item To obtain the intracavity power higher than the critical power, the Kerr medium should be displaced longer than the effective thickness ($d/n_0$) from the center of the concave mirrors. 
\end{itemize}
\section{Conclusion}
We demonstrated the ABCD-matrix analysis of beam propagation in the ring resonator having the peak power close to or much larger than the critical power, $P_\mathrm{c}$.
Our analysis considered the nonlinear propagation caused by self-focusing in the high-peak-power regime.
First, we evaluated the beam spot size as a function of the resonator parameters, $z$ and $x$ in Fig. 1, and the intracavity power, $P$.
When the intracavity power was in the order of $P_\mathrm{c}$, we found a significant dependence on the medium position, $x$.
Moreover, we found that the stable region was continuously deformed with increasing the intracavity power.
The deformation of the stable region indicated that initially stable CW propagation could be evolved into unstable pulse propagation and vice versa, depending on the alignment of the concave mirrors and the Kerr-medium position. 
Second, we evaluated the pulse stability by using the $\delta_K$ parameter expressed by Eq. \ref{deltaK} to find the optimal condition of the resonator from the low- to the high-intracavity-power regime. 
{
These two analyses showed the essential fact that the pulses with the peak power larger than $P_\mathrm{c}$ do not favor the condition of the Kerr medium close to the center of the concave mirrors.
On the other hand, the propagation of these high-peak-power pulses becomes stable by displacing the Kerr medium from the center position.
Therefore, we conclude that stable KLM can be achieved even for ring resonators with $P/P_\mathrm{c}>1$, provided that they are properly aligned according to the above-mentioned guidelines. 
}
%
\section*{Acknowledgement}
This work was supported by JSPS KAKENHI grant numbers 25707024, 26247049, and JP17H06205, by the Photon Frontier Network Program of the Ministry of Education, Culture, Sports, Science and Technology (MEXT), MEXT Quantum Leap Flagship Program (MEXT Q-LEAP) Grant Number JPMXS0118067246. P. X. was supported by Advanced Leading Graduate Course for Photon Science (ALPS). We thank Kazunori Naganuma for his help with the numerical analysis.


\section*{Reference}

\begin{thebibliography}{30}
\bibitem{Spence1991}
D. E. Spence, P. N. Kean, and W. Sibbett, Opt. Lett. {\bf{16}}, 42--44 (1991).
\bibitem{Evans1992}
J. M. Evans, D. E. Spence, and W. Sibbett and B. H. T. Chai and A. Miller, Opt. Lett. {\bf{17}}, 1447--1449 (1992).
\bibitem{Sennaroglu1994}
A. Sennaroglu, C. R. Pollock, and H. Nathel, Opt. Lett. {\bf{19}}, 390--392 (1994).
\bibitem{Petrov1997}
V. Petrov, U. Griebner, D. Ehrt, and W. Seeber, Opt. Lett. {\bf{22}}, 408--410 (1997).
\bibitem{Cizmeciyan2009}
M. N. Cizmeciyan, H. Cankaya, A. Kurt, and A. Sennaroglu, Opt. Lett. {\bf{34}}, 3056--3058 (2009).
\bibitem{Brabec1992}
T. Brabec, Ch. Spielmann, P. F. Curley, and F. Krausz, Opt. Lett. {\bf{17}}, 1292--1294 (1992).
\bibitem{Magni1993}
V. Magni, G. Cerullo, and S. D. Silvestri, Opt. Commun. {\bf{96}}, 348--355 (1993).
\bibitem{Magni1993v2}
V. Magni, G. Cerullo, and S. D. Silvestri, Opt. Commun. {\bf{101}}, 365--370 (1993).
\bibitem{Cerullo1994}
G. Cerullo, S. D. Silvestri, V. Magni, and L. Pallaro, Opt. Lett. {\bf{19}}, 807--809 (1994).
\bibitem{Magni1996}
V. Magni, J. Opt. Soc. Am. B {\bf{13}}, 2498--2507 (1996).
\bibitem{Georgiev1992}
D. Georgiev, J. Herrmann, and U. Stamm, Opt. Commun. {\bf{92}}, 368--375 (1992).
\bibitem{Brabec1993}
T. Brabec, P. F. Curley, Ch. Spielmann, E. Wintner, and A. J. Schmidt, J. Opt. Soc. Am. B {\bf{10}}, 1029--1034 (1993).
\bibitem{Heatley1993}
D. R. Heatley, A. M. Dunlop, and W. J. Firth, Opt. Lett. {\bf{18}}, 170--172 (1993).
\bibitem{Agnesi1994}
A. {Agnesi}, IEEE J. Quantum Electron. {\bf{4}}, 1115--1121 (1994).
\bibitem{Jirauschek2003}
C. Jirauschek, F. X. K\"{a}rtner, and U. Morgner, J. Opt. Soc. Am. B {\bf{20}}, 1356--1368 (2003).
\bibitem{Wright2020}
L. G. Wright, P. Sidorenko, H. Pourbeyram, Z. M. Ziegler, A. Isichenko, B. A. Malomed, C. R. Menyuk, D. N. Christodoulides, and F. W. Wise, Nature Phys. (2020).
\bibitem{Donin2012}
V. I. Donin, D. V. Yakovin, and A. V. Gribanov, Opt. Lett. {\bf{37}}, 338--340 (2012).
\bibitem{Yefet2013}
S. Yefet and A. Pe'er, Appl. Sci. {\bf{3}}, 694--724 (2013).
\bibitem{Klenner2015}
A. Klenner and U. Keller, Opt. Express {\bf{23}}, 8532--8544 (2015).
\bibitem{Li2017}
Z. Li, J. Peng, J. Yao, and M. Han, Opt. Laser Technol. {\bf{23}}, 1--5 (2017).
\bibitem{Sugiyama2018}
N. Sugiyama, H. Tanaka, and F. Kannari, Jpn. J. Appl. Phys. {\bf{57}}, 052701 (2018).
\bibitem{Fibich2000}
G. Fibich and A. L. Gaeta, Opt. Lett. {\bf{25}}, 335--337 (2000).
\bibitem{Larotonda2003}
M. A. Larotonda, Opt. Commun. B {\bf{228}}, 381--388 (2003).
\bibitem{Bartels2002}
A. Bartels and H. Kurz, Opt. Lett. {\bf{27}}, 1839--1841 (2002).
\bibitem{Bartels1999}
A. Bartels, T. Dekorsy, and H. Kurz, Opt. Lett. {\bf{24}}, 996--998 (1999).
\bibitem{Brabec1991}
T. Brabec, Ch. Spielmann, and F. Krausz, Opt. Lett. {\bf{24}}, 1961--1963 (1991).
\bibitem{Kartner1998}
F. X. {K\"{a}rtner}, J. A. {der Au}, and U. {Keller}, IEEE J. Sel. Top. Quantum Electron. {\bf{24}}, 159--168 (1998).
\bibitem{Haus2000}
H. A. {Haus}, IEEE J. Sel. Top. Quantum Electron. {\bf{6}}, 1173--1185 (2000).
\end{thebibliography}
\end{document}